\documentclass[prb,twocolumn,superscriptaddress,nofootinbib]{revtex4-2}

\usepackage{amsmath}
\usepackage{graphicx}
\usepackage{xcolor}
\usepackage{hyperref}
\usepackage{placeins}

\begin{document}

\title{Liquid-gas analog multicriticality in a frustrated Ising bilayer}

\author{Yuchen Fan}
\email{yuchenfan@imu.edu.cn}
\affiliation{Research Center for Quantum Physics and Technologies, School of Physical Science and Technology, Inner Mongolia University, Hohhot 010021,China}

\begin{abstract}
We report the discovery of a novel multicritical point that extends the liquid–gas paradigm to systems with competing symmetry-breaking orders. Using large-scale Monte Carlo simulations of a frustrated bilayer Ising antiferromagnet with tunable couplings, we map out a rich finite-temperature phase diagram hosting three ordered phases separated by both continuous and first-order transitions. By tuning the couplings, a tricritical line and a critical endpoint line converge into a single multicritical line. At all points along the multicritical line, symmetry-distinct phases exhibit identical leading critical behavior---consistent with the tricritical Ising universality class---while the subleading exponent exhibits a sharp shift from $y_g = 0.8$ to $y_g = 1$. This shift reflects an emergent $Z_2$ symmetry akin to that of the liquid–gas critical point, but realized here at a genuine multicritical point involving simultaneous microscopic symmetry breaking. Our results establish a new universality scenario in which emergent symmetry preserves the leading class but reorganizes subleading scaling, providing a general mechanism for symmetry-enforced multicriticality.
\end{abstract}

\maketitle

\section{Introduction}
Symmetry plays a central role in classifying critical phenomena. While conventional phase transitions involve spontaneous breaking of explicit microscopic symmetries, many-body systems can also host emergent symmetries that reshape their critical behavior. A paradigmatic case is the liquid–gas critical point, where the critical equivalence of low- and high-density phases gives rise to an emergent \textit{inter-phase}  \( Z_2 \) symmetry that constrains the effective theory and yields Ising universality~\cite{Cardy1996,Chaikin1995}. Analogous emergent Ising criticality has been observed in quantum magnets, including frustrated bilayer~\cite{Stapmanns2018,Weber2022,Fan2024}, trilayer~\cite{Lukas2022}, and diamond-decorated~\cite{Stre2023,Caci2023} Heisenberg models, as well as in Shastry--Sutherland~\cite{Mila2021,Wang2023} and pyrochlore systems~\cite{Tang2022}, suggesting a broader role for emergent symmetry across classical and quantum systems.

This motivates a fundamental question: what universal structures emerge when such an emergent symmetry coexists with competing symmetry-breaking orders? Such a scenario defines a distinct multicritical point, governed by the interplay between emergent and underlying microscopic symmetries—beyond the conventional liquid–gas paradigm. The emergent symmetry then acquires new significance: it may reshape the renormalization group flow and modify the universality class. As a natural extension of liquid–gas criticality, such multicritical points define a previously unexplored class of critical phenomena, offering a new lens on how emergent symmetries organize critical scaling in statistical systems.

To address this question, we investigate a frustrated bilayer Ising model with tunable intra- and interlayer couplings. This model imposes a local exclusion constraint between symmetry-distinct ordering sectors, and supports an emergent symmetry that is restored at a finite-temperature multicritical point. Unlike its SU(2)-symmetric Heisenberg counterpart, where the Mermin–Wagner theorem forbids finite-temperature ordering and the sign problem limits quantum Monte Carlo simulations to the fully frustrated regime~\cite{Stapmanns2018}, our classical model supports robust thermal transitions and enables the investigation of rich multicritical behavior under widely tunable conditions.

\begin{figure}
\includegraphics[height=9.0cm,width=8.2cm]{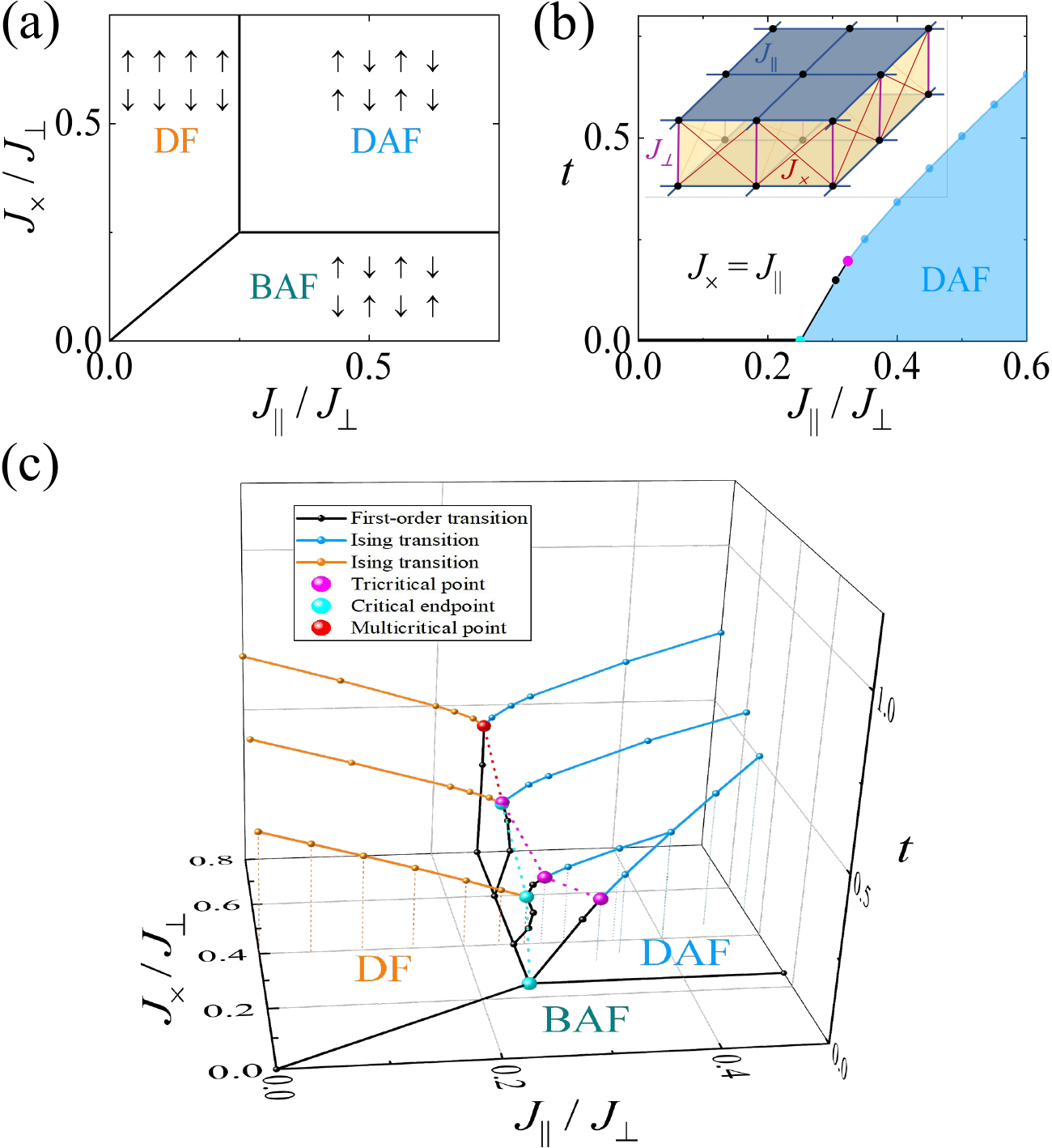}
\caption{
(a) Ground-state phase diagram in the $(J_\parallel / J_\perp, J_\times / J_\perp)$ plane.  
(b) Finite-temperature phase diagram on the $J_\times = J_\parallel$ slice, showing finite temperature first-order (black) and second-order Ising (blue) transitions; their junction defines a tricritical Ising point (magenta). The cyan point at ($t=0$, $J_\parallel/J_\perp = 0.25$) marks the triple point in (a), which also serves as a critical endpoint where the DF transition line can terminate (e.g., on the $J_\times/J_\perp = 0.25$ slice). Inset: bilayer square-lattice geometry and interactions.
(c) 3D phase diagram in $(t, J_\parallel / J_\perp, J_\times / J_\perp)$ space, shown for slices at $J_\times = J_\parallel$ and at fixed $J_\times/J_\perp = 0.4$, $0.6$, and $0.8$. Representative results for $J_\times/J_\perp = 0.4$ and $0.8$ are presented in the main text, while results for the $J_\times = J_\parallel$ slice are given in Appendix~\ref{app:slice1}. Yellow and blue lines denote continuous Ising transitions into the DF and DAF phases; black lines indicate first-order transitions. Magenta and cyan dots mark tricritical points and critical endpoints within each slice, which evolve into tricritical and critical endpoint lines merging into a single multicritical line at strong coupling. The finite-temperature extension of the BAF phase, symmetric to DF, is omitted here for clarity.
}
\label{fig:1}
\end{figure}

The resulting phase diagram [Fig.~\ref{fig:1}(c)] hosts three symmetry-breaking phases—dimer-ferromagnetic (DF), dimer-antiferromagnetic (DAF), and bilayer-antiferromagnetic (BAF). Using large-scale Monte Carlo simulations and field-mixing analysis, we identify a tricritical line and a critical endpoint line that, upon tuning the coupling ratio, converge into a single multicritical line. At all points along the multicritical line, the DF and DAF phases break distinct \( Z_2 \) symmetries yet share identical leading critical behavior consistent with the 2D tricritical Ising universality class. Strikingly, the subleading exponent exhibits an abrupt shift from \( y_g = 0.8 \) to \( y_g = 1 \), indicating that an emergent \( Z_2 \) symmetry—arising from a local exclusion constraint—modifies the renormalization group flow and elevates the associated exponent. Our results establish a new class of multicriticality in statistical systems, where emergent symmetry restructures subleading scaling while preserving the leading universality—extending the liquid–gas paradigm to symmetry-breaking systems.

The rest of the paper is organized as follows. In Sec.~\ref{sec_II} we introduce the model and method. In Sec.~\ref{sec_III} we present the main results, including the ground-state phase diagram, the tricritical behavior and critical endpoint, and the emergence of the multicritical point. Section~\ref{sec_IV} concludes with a discussion.

\section{Model and method}\label{sec_II}
We consider a classical Ising model on a bilayer square lattice, where spins $s_{i,\ell} = \pm 1/2$ reside on each site $i$ of layer $\ell = 1,2$. The Hamiltonian includes vertical interlayer coupling $J_\perp$, intralayer nearest-neighbor coupling $J_\parallel$, and diagonal interlayer coupling $J_\times$, and is given by
\begin{equation}
\mathcal{H} = J_\perp \sum_i s_{i,1} s_{i,2}
+ J_\parallel \sum_{\langle i,j \rangle,\ell} s_{i,\ell} s_{j,\ell}
+ J_\times \sum_{\langle i,j \rangle_\times} s_{i,1} s_{j,2},
\label{eq:H}
\end{equation}
where $\langle i,j \rangle$ denotes intralayer nearest neighbors and $\langle i,j \rangle_\times$ refers to diagonal interlayer (next-nearest neighbor) pairs. This model exhibits a nontrivial duality-like symmetry: exchanging $s_{i,1} \leftrightarrow s_{i,2}$ on one sublattice, together with interchanging $J_\parallel \leftrightarrow J_\times$, leaves the Hamiltonian invariant. This symmetry underlies the symmetric structure of the phase diagram in Fig.~\ref{fig:1}. We rewrite Eq.~\eqref{eq:H} using the dimer-sum and dimer-difference variables \( T_i = s_{i,1} + s_{i,2} \) and \( D_i = s_{i,1} - s_{i,2} \), which recast the bilayer system as a single square lattice of dimers:
\begin{equation}
\mathcal{H} = \frac{J_\perp}{2} \sum_i T_i^2 + \frac{J_\parallel + J_\times}{2} \sum_{\langle i,j \rangle} T_i T_j + \frac{J_\parallel - J_\times}{2} \sum_{\langle i,j \rangle} D_i D_j.
\label{eq:H_comp}
\end{equation}
Here, we have dropped a constant $-J_\perp {L^2}/4$, where ${L^2}$ is the number of square lattice sites. This formulation highlights a formal resemblance to the spin-1 Blume–Capel model when $J_\parallel = J_\times$, with $T_i$ playing the role of an effective three-level variable; however, the mapping is not exact since the $T_i = 0$ state in our model corresponds to a twofold degeneracy.  Quantum versions of this model have attracted considerable interest~\cite{Richter2006,Oleg2010,Richter2018,Oitmaa2021}, with sign-problem-free Monte Carlo simulations in the fully frustrated limit ($J_\times = J_\parallel$) revealing a thermal Ising critical point~\cite{Stapmanns2018}.

Simulations are performed using the Metropolis algorithm with periodic boundary conditions and system sizes up to $L = 48$. We define a reduced temperature \( t = T/J_\perp \) for convenience.
To distinguish between the ordered phases, we compute the static spin structure factor,
$
S(\mathbf{q}) = \frac{1}{2L^2} \sum_{i,\ell; j,\ell'}
\langle s_{i,\ell} s_{j,\ell'} \rangle
\, e^{i \mathbf{q} \cdot (\mathbf{r}_{i,\ell} - \mathbf{r}_{j,\ell'})},
$
which exhibits characteristic peaks at ordering wave vectors \( \mathbf{q}' \) corresponding to each phase.
To probe continuous transitions, we compute the effective correlation length
$
\xi = \left( \frac{L}{2\pi} \right) \sqrt{ \frac{S(\mathbf{q'})}{S(\mathbf{q'} + \delta \mathbf{q})} - 1 },
$
and define the corresponding order parameter as
$
O = \sqrt{ \frac{S(\mathbf{q'})}{2L^2} },
$
where $\delta \mathbf{q}$ is the smallest nonzero wave vector accessible on the finite lattice.
Near a continuous transition, these observables obey finite-size scaling forms
$
\xi = L \, F_{\xi}(|x| L^{1/\nu}), \quad
O = L^{-\beta/\nu} \, F_m(|x| L^{1/\nu}),
$
where $x$ quantifies the proximity to the critical point, and $\nu$, $\beta$ are the correlation length and order parameter exponents, respectively~\cite{Sandvik2010}.

We employ a field-mixing method, previously developed for spin fluids and the Blume–Capel model~\cite{Bruce1992,Wilding1992,Wilding1996,Plascak2013,Kwak2015,Mataragkas2023},
to accurately locate the multicritical point and extract its associated scaling exponents.
Two mixed scaling fields are introduced: \( \lambda = \mu + r \beta \) and \( g = \beta + s \mu \), with \( \beta = J_\perp / (2T) = 1/(2t)\) and
 \( \mu = J_\parallel / (2T) \) under either fixed $J_\times$ or $J_\times=J_\parallel$, where \( r \) and \( s \) are field-mixing parameters.
Here, \( g \) defines the direction tangent to the transition line in the \( (\mu, \beta) \) plane, while \( \lambda \) spans a generic direction.
Determining the tangent field \( g \) (i) allows precise localization of the multicritical point and (ii) isolates the subleading exponent \( y_g \), which is otherwise masked by the leading exponent \( y_t \) along generic directions.
The variable conjugate to \( \lambda \) is proportional to
$Q = (\epsilon - s n),
$
with
$
\left< \epsilon \right> = \frac{1}{L^2} \frac{\partial \ln Z}{\partial \mu},
\left< n \right > = \frac{1}{L^2} \frac{\partial \ln Z}{\partial \beta}
$
~\cite{Kwak2015,Mataragkas2023}.
We construct the centered variable \( \tilde{Q} = Q - \langle Q \rangle \), and locate the finite-size transition point by adjusting \( \mu \) and  \( s \) at fixed \( \beta \) such that the distribution \( P_L(\tilde{Q}) \) exhibits symmetric double peaks.
The multicritical point is located from the size-independent crossing of the fourth-order cumulant along the transition line,
$
U_Q = 1 - \frac{\langle \tilde{Q}^4 \rangle}{3 \langle \tilde{Q}^2 \rangle^2},
$
whose finite-size scaling yields the subleading exponent \( y_g \).
The leading exponent \( y_t \) is extracted by collapsing the distribution according to
$
P_L(\tilde{Q}) = L^{d - y_t} \, \tilde{p}(L^{d - y_t} \tilde{Q})
$
~\cite{Kwak2015,Mataragkas2023}.

\section{Results}\label{sec_III}

In this section we present our numerical results in three steps. First, we establish the zero-temperature phase diagram by comparing the ground-state energies of candidate phases. Second, we examine the finite-temperature behavior at representative coupling ratios, identifying tricritical point and critical endpoint. Finally, we show how these features merge into a multicritical point at stronger coupling, and analyze the associated scaling properties.

\subsection{Ground-state phase diagram}
To determine the zero-temperature phase diagram, we analytically evaluate and compare the ground-state energies of candidate spin configurations across different regions of the coupling space. Specifically, we consider three ordered states: (i) the dimer–ferromagnetic (DF) phase, where each layer is ferromagnetically ordered with opposite spin alignment between the two layers; (ii) the dimer–antiferromagnetic (DAF) phase, where each layer exhibits Néel order while spins across layers are ferromagnetically aligned; and (iii) the bilayer antiferromagnetic (BAF) phase, where each layer exhibits Néel order with opposite spin alignment between the two layers.
The ground-state energy per dimer for each phase is given by:
\begin{align}
E_{\mathrm{GS}}^{\mathrm{DF}} &= J_\parallel - J_\times, \\
E_{\mathrm{GS}}^{\mathrm{DAF}} &= \frac{J_\perp}{2} - J_\parallel - J_\times, \\
E_{\mathrm{GS}}^{\mathrm{BAF}} &= -J_\parallel + J_\times.
\end{align}
By comparing these energies, we identify three first-order phase boundaries:  
(i) DF--DAF with $J_\perp = 4 J_\parallel$,  
(ii) DF--BAF with $J_\times = J_\parallel$, and  
(iii) DAF--BAF with $J_\perp = 4 J_\times$.

These three boundaries intersect at a single three-phase point located at 
\((J_\parallel/J_\perp, J_\times/J_\perp) = (0.25, 0.25)\), where all three states are energetically degenerate. The resulting zero-temperature phase diagram is presented in Fig.~\ref{fig:1}(a). Notably, the DF–BAF boundary corresponds to a line of macroscopically degenerate ground states. The degeneracy arises from competition between intra- and interlayer couplings, which allows a macroscopically large set of local spin configurations to minimize the energy. This extensive degeneracy prevents the system from selecting a unique ordered configuration upon heating, leading to entropy-driven disorder. Consequently, the DF–BAF boundary cannot sustain long-range order at any finite temperature and no symmetry-breaking transition is expected upon heating. Consistent with this picture, Monte Carlo simulations in Fig.~\ref{fig:1}(b) show no thermal ordering along the DF–BAF boundary despite its energetic relevance at zero temperature.

Each of the three ordered phases breaks a distinct combination of the model’s discrete symmetries. All three phases break the global spin-flip symmetry. The DF phase breaks the layer-exchange symmetry, while preserving in-plane translation. The DAF phase preserves layer symmetry but breaks in-plane translational symmetry. The BAF phase breaks both. Additionally, the model also exhibits a duality-like invariance: applying $s_{i,1} \leftrightarrow s_{i,2}$ on one sublattice (e.g., sublattice A), together with exchanging $J_\parallel \leftrightarrow J_\times$, leaves the Hamiltonian invariant. This maps the DF and BAF states onto one another and motivates the choice of $J_\perp$ as the energy unit, enabling a symmetric representation of the phase diagram in the $(J_\parallel/J_\perp, J_\times/J_\perp)$ parameter space. Overall, the zero-temperature phase diagram provides a symmetry-based foundation for understanding how first-order lines extend into the thermal regime, leading to tricritical and critical endpoints discussed below.

\begin{figure}
\includegraphics[width=\columnwidth]{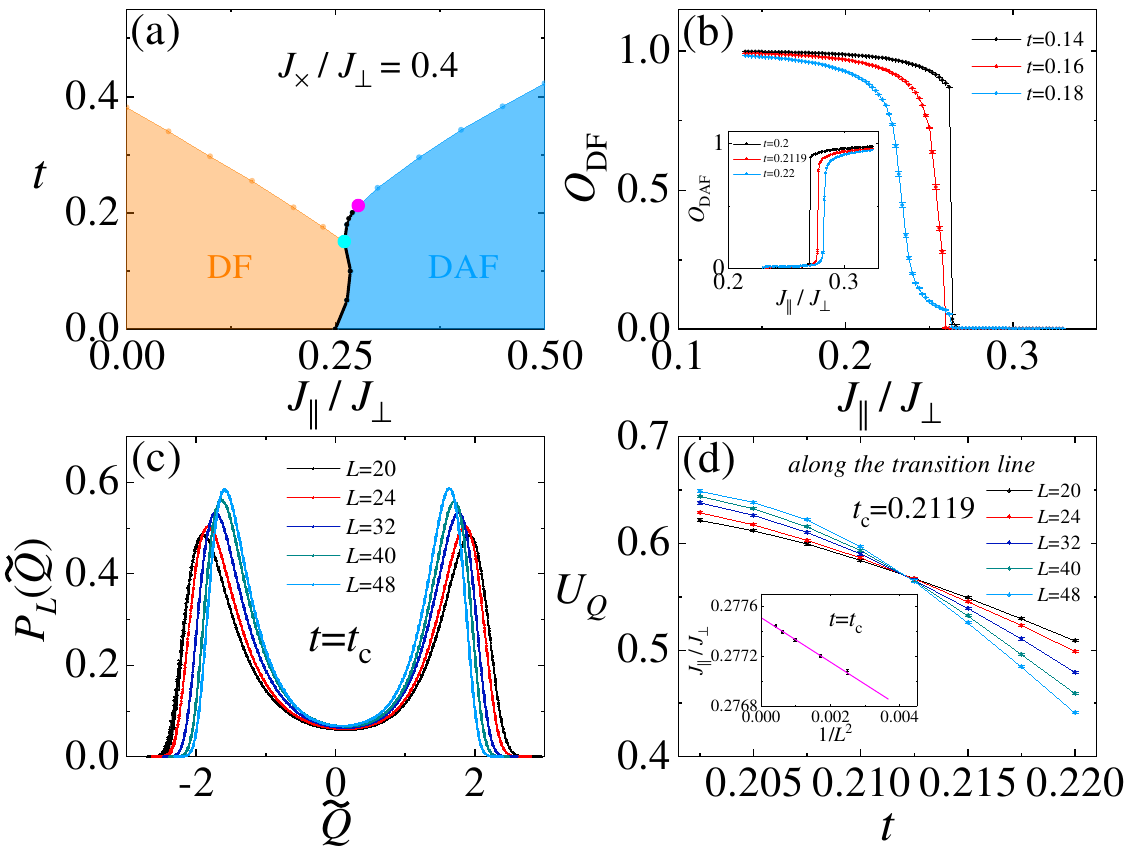}
\caption{
(a) Finite-temperature phase diagram at $J_\times / J_\perp = 0.4$. The black line denotes a first-order transition, with a thermal CEP (cyan) and a TCP (magenta) separating first- and second-order transitions.  
(b) DF order parameter versus $J_\parallel / J_\perp$ at various temperatures near the CEP. Inset: DAF order parameter across the transition line near the TCP.  
(c) Symmetric $P_L(\tilde{Q})$ at $t = t_c$ for various $L$.  
(d) Fourth-order cumulant $U_Q$ along the transition line, determining the tricritical temperature $t_c = 0.2119(2)$. Inset: finite-size extrapolation of the transition points yields $(J_\parallel / J_\perp)_c$=0.27751(3).
}
\label{fig:2}
\end{figure}

\subsection{Tricriticality and critical endpoint}
We begin our analysis of finite-temperature multicritical behavior by fixing $J_\times / J_\perp = 0.4$ and mapping the phase diagram in the $(t, J_\parallel / J_\perp)$ plane [Fig.~\ref{fig:2}(a)]. At low temperatures, two ordered phases—DF and DAF—emerge, separated by a first-order transition line. Along this line, the DF–paramagnetic second-order transition terminates at a thermal critical endpoint (CEP). At this point, critical fluctuations develop only along the direction tangent to the first-order line, while all other directions intersect the first-order surface and lead to discontinuous jumps in the order parameter. As such, the thermal CEP does not represent a multicritical point in the renormalization-group sense.
In contrast, the DAF–paramagnetic transition changes character at higher temperatures: a tricritical point (TCP) separates the low-temperature first-order regime from a high-temperature continuous transition. Finite-size scaling analysis (presented below) confirms that this point lies in the tricritical Ising universality class, as realized in the 2D spin-1 Blume–Capel model.

To characterize the thermal CEP, we track the DF order parameter as a function of $J_\parallel / J_\perp$ at fixed temperatures near the transition. As shown in Fig.~\ref{fig:2}(b), a clear discontinuous jump appears for $t < t_{\mathrm{CEP}} = 0.152(3)$, consistent with a first-order transition. For $t_{\mathrm{CEP}} < t < t_c = 0.2119(2)$, the order parameter initially decreases continuously—indicating a second-order transition—but displays a first-order jump at larger \( J_\parallel / J_\perp \), due to the proximity of the first-order transition line. The inset shows the DAF order parameter, which also exhibits a first-order transition below $t_c$. Above $t_c$, the discontinuity vanishes entirely. Finite-size scaling of the correlation lengths and order parameters confirms that the DF–paramagnetic transition above $t_{\mathrm{CEP}}$ and the DAF–paramagnetic transition above $t_c$ both belong to the 2D Ising universality class (see Appendix~\ref{app:slice2}).

To pinpoint the tricritical point, we employ a field-mixing method and analyze both the probability distribution of the mixed variable and the fourth-order cumulant across system sizes. 
At fixed temperature, we determine the transition point by tuning \( J_\parallel / J_\perp \) and the field-mixing parameter \( s \), and identifying, for each system size \( L \), the coupling where the probability distribution \( P_L(\tilde{Q}) \) exhibits symmetric two-peak coexistence. 
As an illustrative example, Figure~\ref{fig:2}(c) shows the resulting symmetric distributions at the estimated tricritical temperature \( t = t_c \), from which we extract the finite-size transition points, plotted in the inset of Fig.~\ref{fig:2}(d). 
We then evaluate the \( U_L(Q) \) at these transition points and scan the temperature along the transition line. 
As shown in Fig.~\ref{fig:2}(d), the crossing of $U_L(Q)$ for different system sizes determines the tricritical temperature \( t_c \). Fixing this value, we extract the size-dependent transition couplings from Fig.~\ref{fig:2}(c) and extrapolate them to the thermodynamic limit, yielding the tricritical coupling \( (J_\parallel / J_\perp)_c \), as shown in the inset of Fig.~\ref{fig:2}(d).

\begin{figure}
\includegraphics[width=\columnwidth]{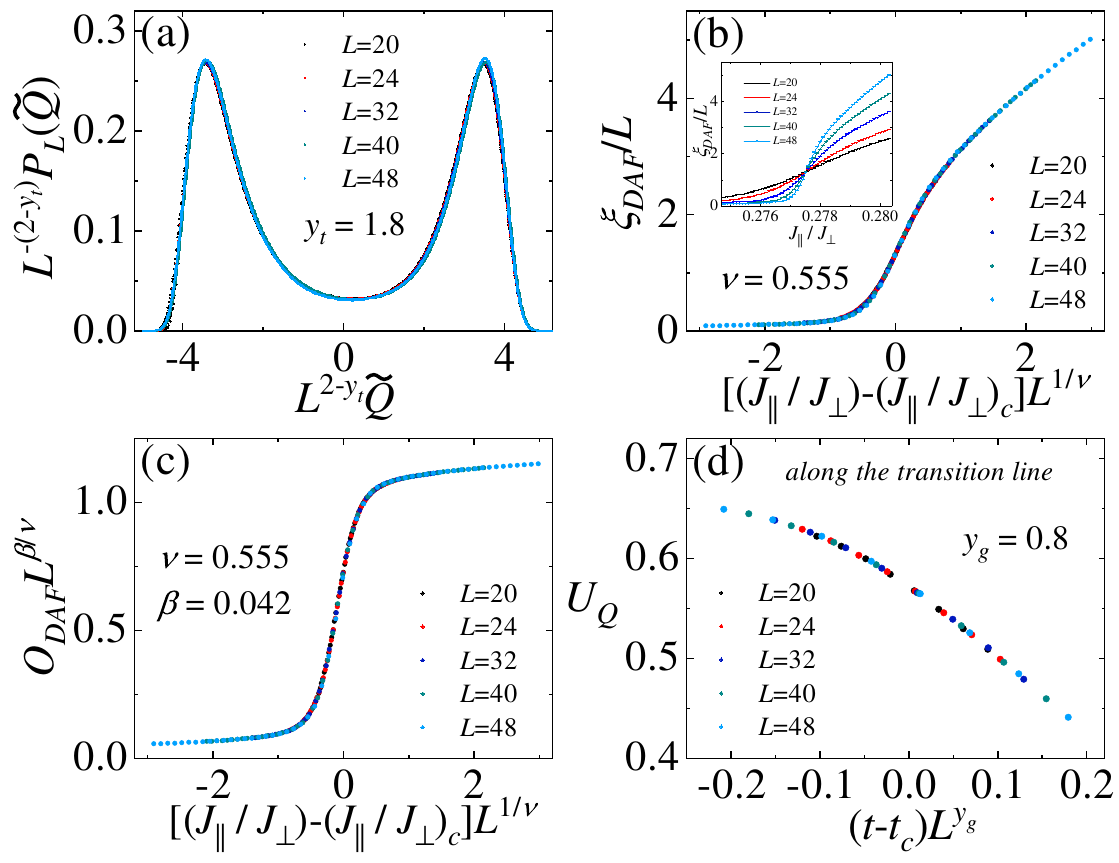}
\caption{Tricritical scaling at $J_\times / J_\perp = 0.4$.
(a) Scaling collapse of the symmetric $P_L(\tilde{Q})$ at $t = t_c$, yielding the leading exponent $y_t = 1.80(2)$.
(b) Scaling collapse of the DAF correlation length at $t = t_c$, giving $\nu = 0.555(6) = 1/y_t$. Inset: Correlation length crossing yields the critical coupling $(J_\parallel / J_\perp)_c = 0.27751(2)$, consistent with the extrapolated value from Fig.~2(d).
(c) Scaling collapse of the DAF order parameter at $t = t_c$, giving $\beta = 0.042(8)$.
(d) Scaling collapse of the fourth-order cumulant $U_Q$ along the DAF transition line, yielding the subleading exponent $y_g = 0.80(2)$.}
\label{fig:3}
\end{figure}

We determine the universality class of the tricritical point via finite-size scaling analysis, combining field-mixing with thermodynamic observables. 
The symmetric $P_L(\tilde{Q})$ collapses according to the scaling form $P_L(\tilde{Q}) = L^{d - y_t} \tilde{p}(L^{d - y_t} \tilde{Q})$, as shown in Fig.~\ref{fig:3}(a), yielding a leading tricritical exponent $y_t = 1.8$, in excellent agreement with the 2D spin-1 Blume--Capel model~\cite{Kwak2015}.
The correlation-length collapse in Fig.~\ref{fig:3}(b) further supports the scaling relation $\nu = 1 / y_t$, while the scaling of the DAF order parameter in Fig.~\ref{fig:3}(c) yields an exponent $\beta = 0.042$, also consistent with tricritical Ising behavior~\cite{Kwak2015}. Additional evidence is provided by the \( U_Q \) along the DAF transition line near the tricritical point, which collapses under the scaling form \( U_Q = \tilde{u}(L^{y_g} (t - t_c)) \), yielding the subleading exponent \( y_g = 0.8 \) [Fig.~\ref{fig:3}(d)]. Taken together, these results firmly establish that the tricritical point at \( J_\times / J_\perp = 0.4 \) belongs to the 2D tricritical Ising universality class, as realized in the spin-1 Blume–Capel model~\cite{Kwak2015}.
This robust scaling behavior reflects the system’s composite structure: the dimer-sum \( T \) sector governs symmetry breaking, while the dimer-difference \( D \) sector remains irrelevant near tricriticality.

\begin{figure}
\includegraphics[width=\columnwidth]{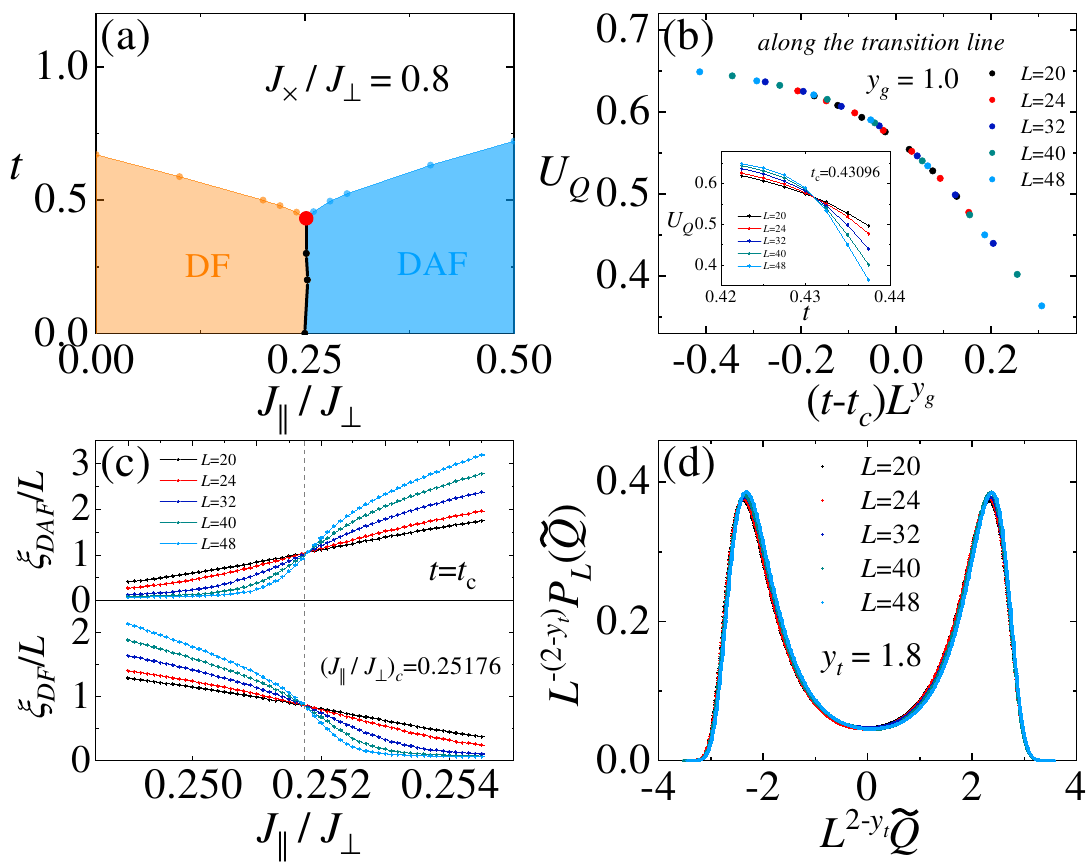}
\caption{
(a) Finite-temperature phase diagram at $J_\times / J_\perp = 0.8$.
(b) Scaling collapse of the fourth-order cumulant $U_Q$ along the transition line yields the subleading exponent $y_g = 1.00(4)$. Inset: crossing of $U_Q$ at $t_c = 0.4309(2)$.
(c) Correlation length crossings for DF and DAF at $t = t_c$ occur at the same coupling $(J_\parallel / J_\perp)_c$=0.25176(4), confirming the convergence of the CEP and TCP.
(d) Scaling collapse of the symmetric $P_L(\tilde{Q})$ at $t = t_c$ gives the leading exponent $y_t = 1.80(2)$, with \( Q = \epsilon \) under optimal field mixing \( s = 0 \).
}
\label{fig:4}
\end{figure}

\subsection{Multicriticality}
At stronger diagonal coupling \( J_\times / J_\perp = 0.8 \), the finite-temperature phase diagram again features DF and DAF ordered phases, separated by a first-order transition line [Fig.~\ref{fig:4}(a)]. In contrast to the \( J_\times / J_\perp = 0.4 \) case, the TCP and CEP now coincide, forming a single multicritical point that terminates the first-order line and connects directly to two continuous transition lines.  
The multicritical temperature \( t_c = 0.4309 \) is determined from the crossing of the \( U_Q \) evaluated along the transition line [inset of Fig.~\ref{fig:4}(b)].  
To verify that the DF and DAF transitions converge at this point, we compare their correlation lengths at \( t = t_c \). Both exhibit crossing behavior at the same coupling ratio [Fig.~\ref{fig:4}(c)], indicating that the two transitions coincide.

The scaling collapse of \( U_Q \) [Fig.~\ref{fig:4}(b)] yields the subleading exponent \( y_g = 1 \), deviating from the tricritical Ising value \( y_g = 0.8 \) but consistent with the standard 2D Ising universality class, signaling a restructuring of the scaling behavior at the multicritical point.  
This shift arises from the competition between the \( T^2 = 1 \) and \( D^2 = 1 \) sectors, subject to the local exclusion constraint \( T_i^2 + D_i^2 = 1 \), which introduces an emergent binary degree of freedom that selects between the DF and DAF orders.  
Under thermal fluctuations, this binary choice gives rise to an emergent \( Z_2 \) symmetry---analogous to the low- and high-density symmetry of the liquid–gas critical point---which is not present microscopically but becomes restored at the multicritical point.
Supporting this, the probability distribution \( P_L(\tilde{Q}) \) becomes symmetric about \( Q = \epsilon \) under optimal field mixing \( s = 0 \) [Fig.~\ref{fig:4}(d), see also Appendix~\ref{app:slice3}], indicating statistical equivalence between DF and DAF configurations at the multicritical point.

\begin{figure}
\includegraphics[width=\columnwidth]{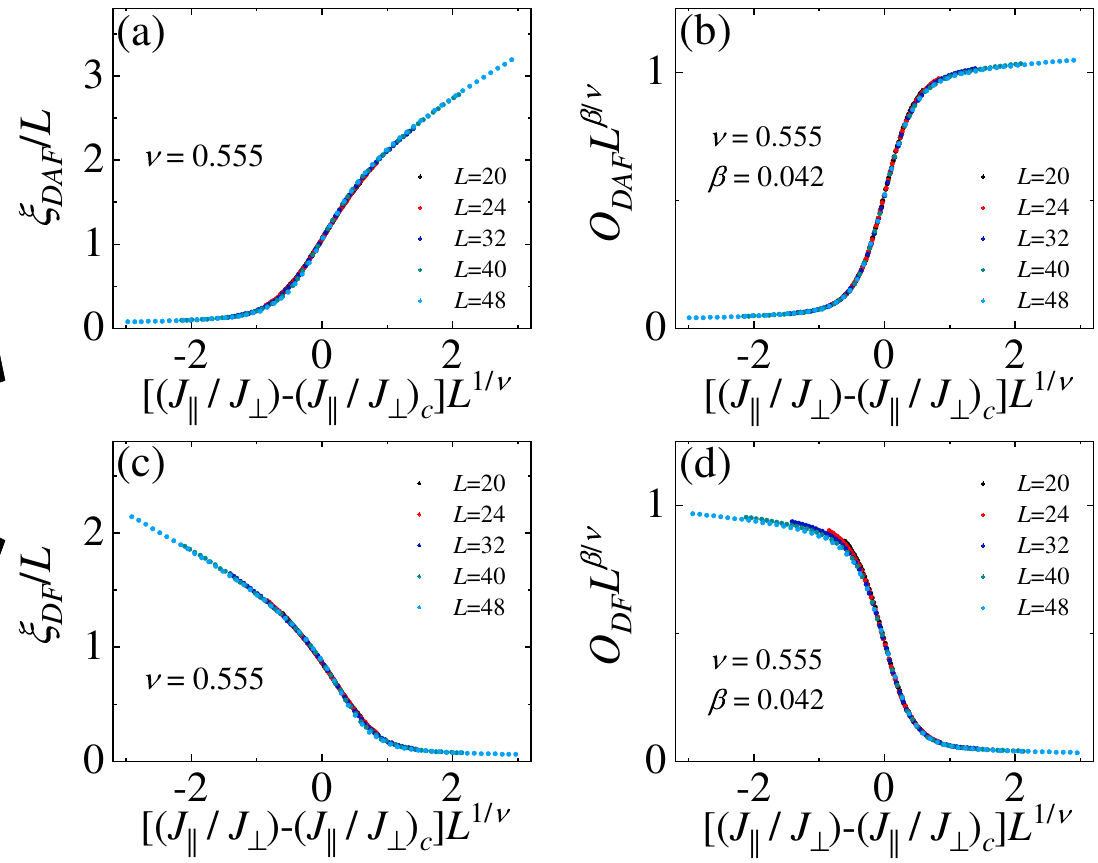}
\caption{
Finite-size scaling collapses at the multicritical point of the $J_\times/J_\perp = 0.8$ slice, where $t=t_c$: correlation lengths for DAF (a) and DF (c) yield a common exponent $\nu = 0.555(6)$, while the corresponding order parameters in (b) and (d) collapse with $\beta = 0.042(9)$.
}
\label{fig:5}
\end{figure}

While the subleading scaling shifts, the leading behavior at the multicritical point exhibits tricritical Ising characteristics. Scaling collapses of the symmetric \( P_L(\tilde{Q}) \) yield a leading thermal exponent \( y_t = 1.80 \) [Fig.~\ref{fig:4}(d)], and the correlation lengths and order parameters for both DF and DAF sectors scale with exponents \( \nu = 0.555 \) and \( \beta = 0.042 \)[Fig.~\ref{fig:5}], respectively, consistent with the tricritical Ising universality class. This behavior naturally arises from the local constraint, under which both the DF and DAF sectors exhibit a combination of continuous and first-order transitions, analogous to those in the spin-1 Blume--Capel model, thereby potentially enabling tricritical Ising scaling~\cite{Cardy1996}.
This complements the effect of the emergent symmetry, which modifies the subleading behavior while preserving the leading scaling.
Identical multicritical behavior is also observed on the $J_\times / J_\perp = 2.0$ slice of the phase diagram (see Appendix~\ref{app:slice4}).
It is worth emphasizing that the critical behavior exhibited by the DF sector (which would not occur at a mere CEP), the abrupt change of the subleading exponent, and the vanishing of the mixing parameter $s=0$ at the multicritical point together provide numerical evidence that the TCP and CEP genuinely merge into a multicritical point, rather than remaining in a scenario where they only approach each other closely. We also note that this merging implies that an initial convergence must occur within a finite range of parameters, but identifying its precise location is not the main focus here.

\section{Discussion}\label{sec_IV}
Our study reveals a unifying mechanism for emergent \( Z_2 \) symmetry in both classical and quantum frustrated bilayer systems, rooted in local constraints that enforce mutual exclusivity between competing orders. In the fully frustrated bilayer Heisenberg model, exact spin conservation on each dimer separates singlet and triplet sectors, giving rise to a binary degree of freedom that supports emergent Ising criticality~\cite{Stapmanns2018}. In our classical bilayer Ising model, an analogous local exclusion arises between the \( T \) and \( D \) sectors, which correspond to symmetry-distinct ordering channels. A key distinction lies in the nature of symmetry breaking: the classical model supports discrete symmetry-breaking transitions, enabling the emergence of TCPs and CEPs. Crucially, the emergent \( Z_2 \) symmetry appears only at the multicritical point, where the DF and DAF phases become equally competitive, thus generalizing the liquid–gas critical point to a setting with spontaneous symmetry breaking.

Our findings establish a new class of multicriticality in a frustrated bilayer Ising model, arising from the interplay between spontaneous symmetry breaking and emergent symmetry. This multicritical structure provides a rare and well-controlled setting where an emergent symmetry modifies subleading critical behavior while preserving the leading exponent. Though rooted in a purely classical constraint structure, the role of the emergent symmetry could provide insight into general organizing principles, potentially shedding light on quantum multicriticality, such as in deconfined quantum critical points~\cite{Hui2016,Bowen2020,Chester2024,Sandvik2024}. Beyond theoretical implications, the underlying mechanism could motivate experimental exploration in programmable platforms such as Rydberg atom arrays and artificial spin-ice systems.

\begin{acknowledgments}
We thank Yuan Wan for insightful suggestions and constructive criticism. We are grateful to Ning Xi, Changle Liu, Bruce Normand, and Rong Yu for foundational collaborations on the study of frustrated bilayer systems. This work was supported by the National Natural Science Foundation of China
(Grant No. 12404176).
\end{acknowledgments}

\section*{Data Availability}
The data that support the findings of this study are openly available~\cite{dataset}.

\onecolumngrid

\appendix 

\section{Tricriticality on the $J_\times = J_\parallel$ slice}\label{app:slice1}
\begin{figure*}[!ht]
\centering
\includegraphics[height=8.8cm,width=16.2cm]{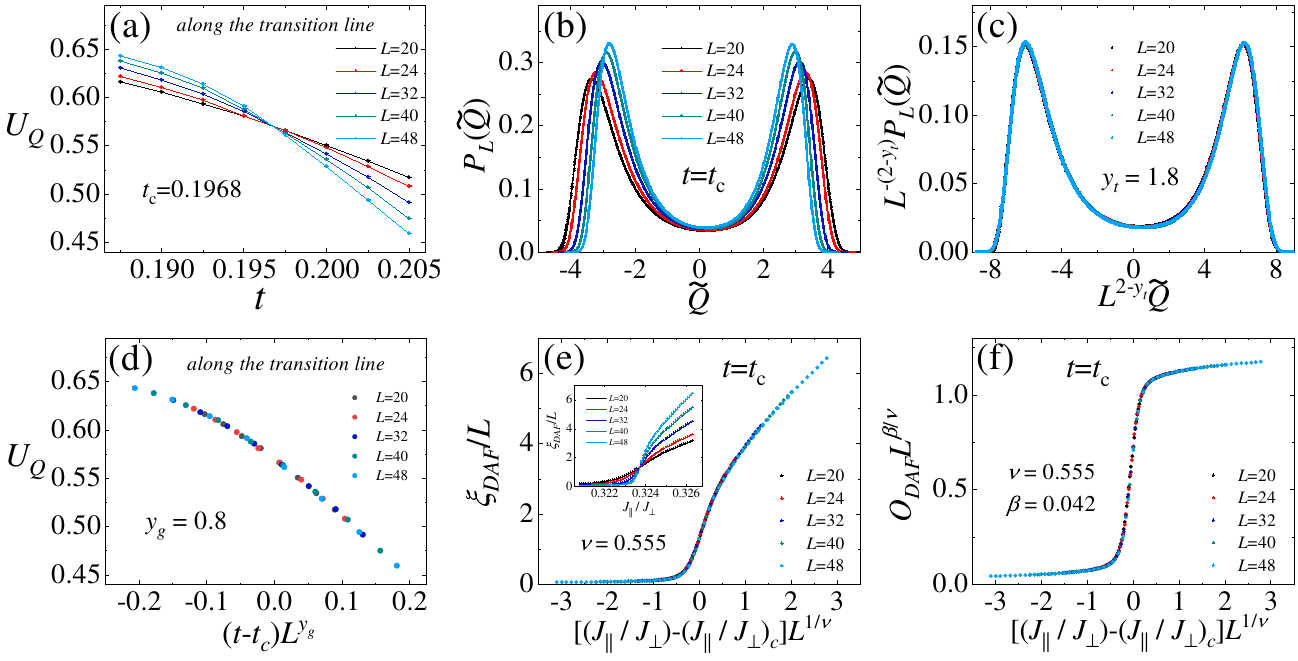}
\caption{(a) Fourth-order cumulant $U_Q$ along the transition line, determining the tricritical temperature $t_c = 0.1968(2)$.
(b) Symmetric $P_L(\tilde{Q})$ at $t = t_c$ for various $L$.
(c) Scaling collapse of the symmetric $P_L(\tilde{Q})$ at $t = t_c$, yielding the leading exponent $y_t = 1.80(2)$.
(d) Scaling collapse of the $U_Q$ along the DAF transition line, yielding the subleading exponent $y_g = 0.80(2)$.
(e) Scaling collapse of the DAF correlation length at $t = t_c$, giving $\nu = 0.555(5) = 1/y_t$. Inset: Correlation length crossing yields the critical coupling $(J_\parallel / J_\perp)_c = 0.32371(2)$.
(f) Scaling collapse of the DAF order parameter at $t = t_c$, giving $\beta = 0.042(8)$.}
\label{fig:S1}
\end{figure*}

\clearpage
\section{Ising transition on the $J_\times / J_\perp = 0.4$ slice}\label{app:slice2}
\vspace{-0.1em}
\begin{figure*}[!ht]  
  \includegraphics[height=8.3cm,width=11.0cm]{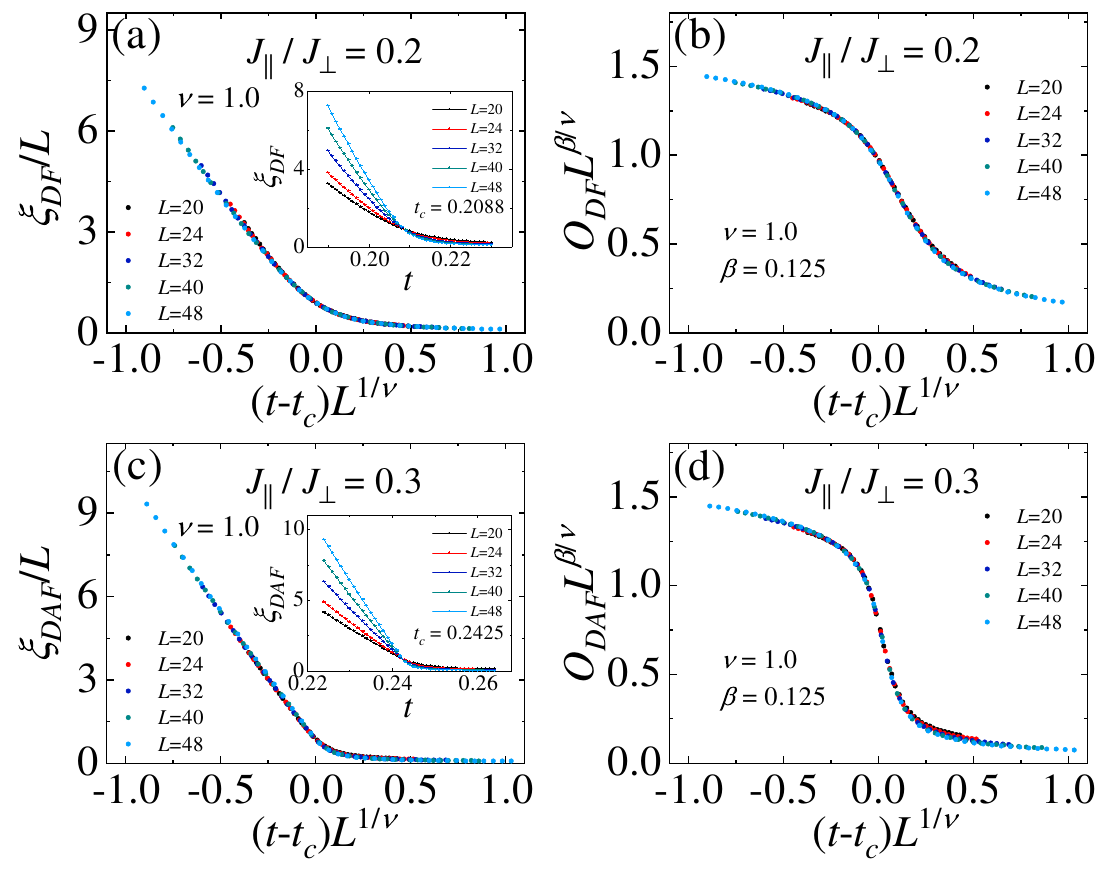}
  \caption{Finite-temperature Ising transitions into the DF and DAF phases at fixed $J_\times / J_\perp = 0.4$. Panels (a, b) correspond to $J_\parallel / J_\perp = 0.2$, deep in the DF regime, and panels (c, d) correspond to $J_\parallel / J_\perp = 0.3$, deep in the DAF regime. (a) and (c) show the finite-size scaling of the correlation length, whose insets exhibit clear crossings at the critical temperature $t_c$. (b) and (d) present the finite-size scaling of the corresponding order parameters for the DF and DAF phases, respectively. All extracted exponents are consistent with standard two-dimensional Ising critical behavior, confirming that continuous thermal transitions into both phases belong to the 2D Ising universality class.}
  \label{fig:S2}
\end{figure*}

\section{Symmetric $P_L(\tilde{Q})$ on the $J_\times / J_\perp = 0.8$ slice}\label{app:slice3}
\vspace{-0.1em}
\begin{figure*}[!ht]  
  \includegraphics[height=5.2cm,width=7.0cm]{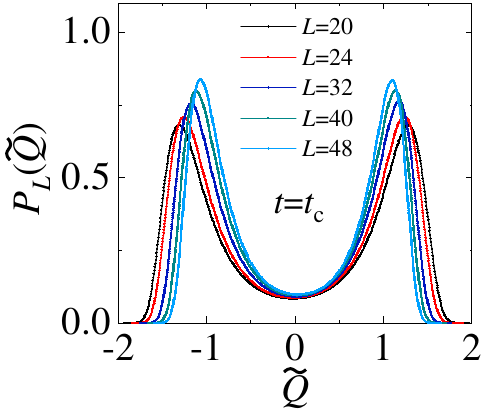}
  \caption{Symmetric $P_L(\tilde{Q})$ at $t = t_c$ for various $L$, with $Q = \epsilon$ under optimal field mixing $s = 0$. This reflects statistical equivalence between DF and DAF configurations, consistent with the emergent $Z_2$ symmetry identified in our scaling analysis.}
  \label{fig:S3}
\end{figure*}

\clearpage
\section{Multicriticality on the $J_\times / J_\perp = 2.0$ slice}\label{app:slice4}
\vspace{-0.1em}
\begin{figure*}[!ht]
\centering
\includegraphics[height=8.8cm,width=16.2cm]{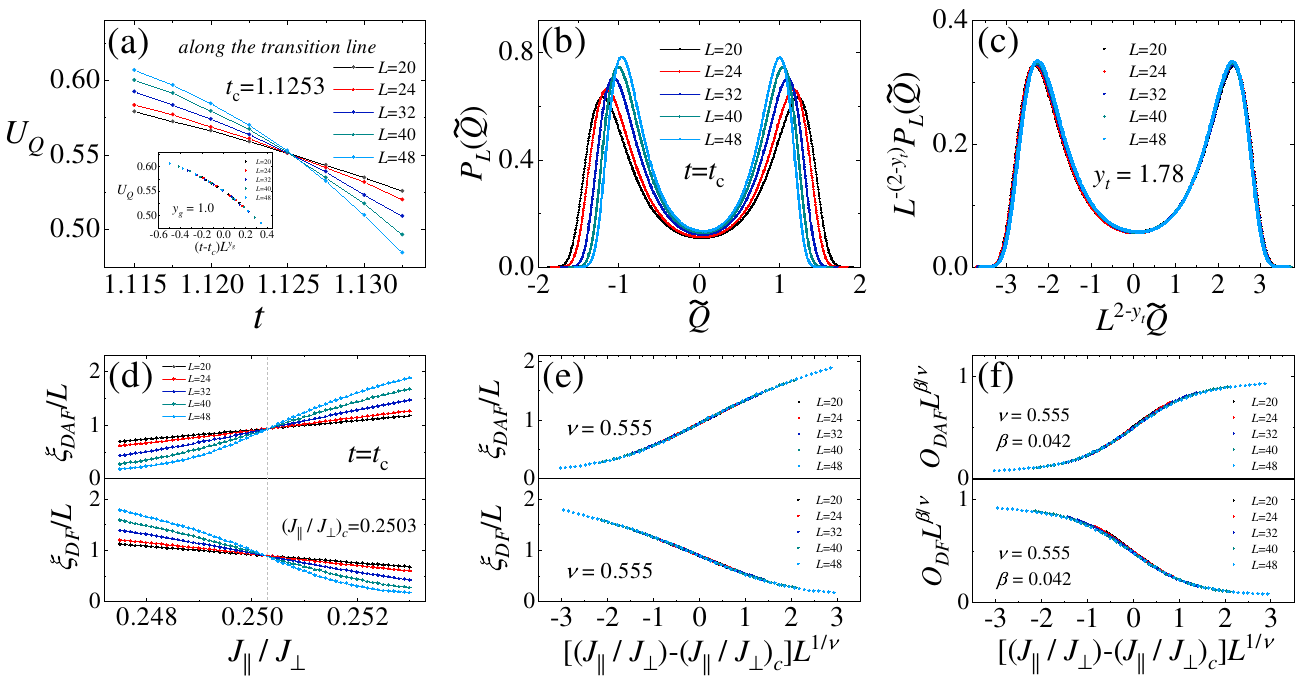}
\caption{(a) Fourth-order cumulant $U_Q$ along the transition line, determining the tricritical temperature $t_c = 1.1253(2)$. Inset: Scaling collapse of the $U_Q$ along the transition line, yielding the subleading exponent $y_g = 1.00(5)$.
(b) Symmetric $P_L(\tilde{Q})$ at $t = t_c$ for various $L$, with $Q = \epsilon$ under optimal field mixing $s = 0$.
(c) Scaling collapse of the symmetric $P_L(\tilde{Q})$ at $t = t_c$, yielding the leading exponent $y_t = 1.78(2)$.
(d) Correlation length crossings for DF and DAF at $t = t_c$ occur at the same coupling $(J_\parallel / J_\perp)_c$=0.25030(6), confirming the convergence of the CEP and TCP.
(e) Scaling collapse of the DF and DAF correlation lengths at $t = t_c$, yielding the common exponent $\nu = 0.555(5)$.
(f) Scaling collapse of the corresponding order parameters, yielding $\beta = 0.042(8)$.}
\label{fig:S4}
\end{figure*}
\twocolumngrid

\bibliography{IsingFB}

\end{document}